\documentclass[prb,reprint,amsmath,amssymb,floatfix,superscriptaddress]{revtex4-1}
\usepackage{graphicx}
\usepackage{hyperref}
\usepackage{enumerate}
\usepackage{bbold}
\usepackage{xcolor}
\renewcommand{\vec}[1]{\mathbf{#1}}
\begin{document}

\title{Spontaneous Edge Current in Higher Chirality Superconductors}

\author{Xin Wang}
\affiliation{Department of Physics and Astronomy, McMaster University, Hamilton, Ontario, L8S 4M1, Canada}
\author{Zhiqiang Wang}
\affiliation{Department of Physics and Astronomy, McMaster University, Hamilton, Ontario, L8S 4M1, Canada}
\author{Catherine Kallin}
\affiliation{Department of Physics and Astronomy, McMaster University, Hamilton, Ontario, L8S 4M1, Canada}
\affiliation{Canadian Institute for Advanced Research, Toronto, Ontario M5G 1Z8, Canada}
\date{\today}

\begin{abstract}
The effects of finite temperature, Meissner screening and surface roughness on the spontaneous edge current
for higher chirality quasi-two dimensional superconductors are studied in the continuum limit using the quasiclassical Eilenberger equations.
We find that the total spontaneous current is non-zero at finite temperature $T$ and maximized near $T=T_c/2$, where $T_c$ is the transition temperature, although it vanishes at $T=0$.
In the presence of surface roughness, we observe a surface current inversion in the chiral $d$-wave case that can be understood in terms of a disorder induced $s$-wave pairing component in the rough surface regime.
This conclusion is supported by a Ginzburg-Landau analysis.
However, this current inversion is non-universal beyond the continuum limit as demonstrated by self-consistent lattice Bogoliubov-de Gennes calculations.
\end{abstract}


\maketitle

\section{Introduction}
Chiral superconductors spontaneously break time reversal symmetry and support chiral Majorana edge modes~\cite{Beenakker2013, Kallin2012, Kallin2016}.
As a consequence, there are spontaneous supercurrents generated at edges. Although the number of chiral edge modes is protected by topology~\cite{Qi2011},
the edge currents are not topologically protected and can strongly depend on microscopic details~\cite{Huang2014, Huang2015, Tada2015, Volovik2015},
since charge is not conserved in a superconducting state, in contrast to a quantum Hall state.

Edge currents, as well as the related total orbital angular momentum of Cooper pairs, have been studied extensively for chiral $p$-wave superconductors~\cite{Huang2014, Huang2015, Tada2015, Volovik2015, Scaffidi2015}.
The major motivation is to reconcile the theoretical prediction of a large edge current~\cite{Matsumoto1999} with the null result observed in scanning probe measurement on $\mathrm{Sr_2RuO_4}$~\cite{Kirtley2007, Hicks2010, Curran2014},
which is believed to be a chiral $p$-wave superconductor~\cite{Kallin2012,Kallin2016,Mackenzie2017}.
Theoretical studies have shown that, in the absence of Meissner screening, the integrated current is substantial at $T=0$~\cite{Ishikawa1977, Kita1998, Stone2004, Sauls2011, Huang2015} and decreases rapidly as $T$ increases~\cite{Kita1998, Sauls2011}.
Previous studies~\cite{Huang2014, Tada2015, Volovik2015} have shown that spontaneous supercurrents in higher chirality superconductors are different from the chiral $p$-wave case.
In particular, the integrated edge currents of the higher chirality superconductors vanish in the semiclassical continuum limit, in stark contrast to the chiral $p$-wave case~\cite{Huang2014}.
These studies focused on $T=0$ and neglected Meissner screening.
More recently, Ref.~\onlinecite{Suzuki2016} studied finite temperature and screening effects on edge currents for higher chirality superconductors in a mesoscopic system (a very narrow cylinder).

In this work, we generalize the study in Ref.~\onlinecite{Suzuki2016} to a half-infinite system, where these effects can be separated from finite size effects
and also examine more closely the explanation for and robustness of current inversion due to disorder.
Following Ref.~\onlinecite{Matsumoto1999, Suzuki2016}, we study the edge currents in the continuum limit for a quasi-two-dimensional chiral superconductor using the quasiclassical Eilenberger equations~\cite{Eilenberger1968}.
Interestingly, we find that, without Meissner screening, the integrated edge currents for higher chirality superconductors are non-zero at finite $T$, unlike at $T=0$,
although they are much smaller than the current of the chiral $p$-wave case.
This finite temperature current is a consequence of the superconducting order parameter variations near a surface.

In Refs.~\onlinecite{Ashby2009, Nagato2011, Lederer2014, Bakurskiy2017}, it has been shown that surface roughness, together with band structure effects, can lead to substantial suppression of the edge current in a chiral $p$-wave superconductor and potentially account for the null result of edge currents in $\mathrm{Sr_2RuO_4}$ experiments.
Here we study rough surface effects on higher chirality superconductors by introducing an impurity self-energy in the quasiclassical Green's function.
As in Ref.~\onlinecite{Suzuki2016}, we find the edge current direction is reversed due to strong surface roughness for chiral $d$-wave pairing in the continuum limit.
However, our calculations together with a Ginzburg-Landau (GL) analysis, suggest a physically more transparent explanation for the current inversion.
We ascribe the inversion to a strong disorder induced sub-dominant $s$-wave component near the interface between the rough surface regime and the bulk.
Near the interface the original $d_{x^2-y^2}$ and $i d_{xy}$ components have almost identical spatial variation, due to the surface and disorder, and their contribution to the current is almost zero.
On the other hand, the induced $s$-wave component is real and can combine with the $i d_{xy}$ component to give a sizable current near the interface if the $s$-wave channel is not too repulsive.
The current resulting from the $s+id_{xy}$ pairing is opposite in direction to that near the specular surface.

However, the current inversion is non-universal beyond the continuum limit.
This is supported by self-consistent lattice Bogoliubov-de Gennes (BdG) calculations, where we explicitly show that the existence of the current inversion depends on edge orientation and chemical potential or band structure effects.

The rest of the paper is organized as follows: in Sec. ~\ref{section:Eilenberger} we outline the self-consistent Eilenberger formalism.
Then we present our results of the edge currents for a specular surface without and with Meissner screening in Sec.~\ref{section:A0} and \ref{section:A}, respectively.
Sec.~\ref{section:R} contains the results with surface roughness. Sec.~\ref{sec:conclusion} contains the discussion and conclusions.

\section{Formalism}
\label{section:Eilenberger}
We consider a semi-infinite ($x>0$) quasi-$2d$ superconductor with a cylindrical Fermi surface independent of $k_z$. The system is described by the following Eilenberger equation~\cite{Eilenberger1968},
\begin{equation}
-iv_{F_x}\frac{d}{dx}\hat{g}(\theta_{\vec{k}},x;\omega_n)=\left[\hat{H}(\theta_{\vec{k}},x;\omega_n),\hat{g}(\theta_{\vec{k}},x;\omega_n)\right],
\label{eq:Eilenberger}
\end{equation}
valid if the characteristic length scale considered is much longer than the Fermi wavelength.
Here, $\hat{g}(\theta_{\vec{k}}, x; \omega_n)$ is the quasi-classical Green's function,
$\theta_{\vec{k}}$ is defined by the direction of the quasiparticle momentum, $\vec{k}=k_F(\cos\theta_{\vec{k}},\sin\theta_{\vec{k}})$, and
$\omega_n=(2n+1)\pi T$ is the Matsubara frequency. The Green's functions does not depend on the magnitude of the momentum $\vec{k}$ as all high energy information involving
$|\vec{k}|\ne k_F$ has been integrated out. Furthermore, since there is no variation along $z$, this coordinate is not shown in $\hat{g}$. As usual, in the Nambu particle-hole space, $\hat{g}(\theta_{\vec{k}}, x; \omega_n)$
is a $2\times 2$ matrix
\begin{equation}
\hat{g}(\theta_{\vec{k}}, x; \omega_n) =
\left(
\begin{array}{cc}
g(\theta_{\vec{k}}, x; \omega_n) & if(\theta_{\vec{k}}, x; \omega_n)\\
-i\bar{f}(\theta_{\vec{k}}, x; \omega_n) & -g(\theta_{\vec{k}}, x; \omega_n)
\end{array}
\right),
\label{eq:Green}
\end{equation}
where $g$ and $f$ are the normal and anomalous parts, respectively. The two components satisfy the normalization relation $ g^2(\theta_{\vec{k}}, x; \omega_n)+f(\theta_{\vec{k}}, x; \omega_n)\bar{f}(\theta_{\vec{k}}, x; \omega_n)=1$, which is a consequence of
$\hat{g}^2(\theta_{\vec{k}}, x; \omega_n)$ being a position independent constant along the quasiparticle trajectory within the Eilenberger quasiclassical
formalism. On the right hand side of Eq.~\eqref{eq:Eilenberger}, $\hat{H}(\theta_{\vec{k}},x;\omega_n)$ is given by
\begin{equation} \label{eq:Hmat}
\hat{H}(\theta_{\vec{k}},x;\omega_n)=
\left(\begin{array}{cc}
i\omega_n-ev_{F_y}A_y(x) & -\Delta(\theta_{\vec{k}}, x)\\
\Delta^{*}(\theta_{\vec{k}}, x) & -i\omega_n+ev_{F_y}A_y(x)
\end{array}\right),
\end{equation}
where, $A_y(x)$ is the y-component of the vector potential satisfying $\nabla \times \vec{A}=B_z(x) \hat{z}$.
$B_z(x)$ is the local magnetic field which can be either generated by the spontaneous edge current or applied externally.
Here, we only consider the spontaneous field and a gauge is chosen such that $A_x(x)\equiv 0$.
The Fermi velocities in Eqs.~\eqref{eq:Eilenberger} and \eqref{eq:Hmat} are defined as $\vec{v}_F=(v_{Fx},v_{Fy})=v_F(\cos\theta_{\vec{k}},\sin\theta_{\vec{k}})$.

The off-diagonal component $\Delta(\theta_{\vec{k}},x)$ in Eq.~\eqref{eq:Hmat} is the chiral superconducting order parameter.
For chiral $m$-wave pairing, it is given by $\Delta(\theta_{\vec{k}},x) = \Delta_1(x)\cos(m \, \theta_{\vec{k}})+\Delta_2(x)\sin(m \, \theta_{\vec{k}})$.
We choose $\Delta_1(x)$ to be real and $\Delta_2(x)$ to be purely imaginary in the bulk such that the order parameter is chiral. $\Delta_1$ and $\Delta_2$ are determined self consistently from the following gap equations:
\begin{subequations}
\begin{align}
\Delta_{1}(x) &= \pi T N_F V \sum_{|\omega_n|<\omega_c}\langle 2\cos(m\theta_{\vec{k}})f(\theta_{\vec{k}},x;\omega_n)\rangle,\\
\Delta_{2}(x) &= \pi T N_F V \sum_{|\omega_n|<\omega_c}\langle 2\sin(m\theta_{\vec{k}})f(\theta_{\vec{k}},x;\omega_n)\rangle.
\end{align}
\label{eq:Gap}
\end{subequations}
$\langle \dots \rangle = \frac{1}{2\pi}\int_{-\pi}^{\pi}d\theta_{\vec{k}}(\dots)$; $\omega_c$ is the pairing energy cutoff; $N_F$ is the normal state density of states at the Fermi energy;  and $V$ is the pairing interaction strength.
The dimensionless attractive interaction strength $N_FV$ is connected to the  superconducting transition temperature $T_c$ by
\begin{equation}
\frac{1}{N_F V} = \log\frac{T}{T_c}+\sum_{n,|\omega_n|\le \omega_c} \frac{1}{n-1/2},
\label{eq:coupling}
\end{equation}
which becomes $T$ independent in the weak coupling limit $T\le T_c\ll \omega_c$.
We will use this equation for $N_F V$ in terms of $T_c$ and $\omega_c$ and rescale all energy quantities by $T_c$. In this way, we do not need to explicitly specify the value of $N_F V$.

We use the Riccati parametrization~\cite{Schopohl1998} to solve for the Green's function matrix, $\hat{g}$. The current density $J_y(x)$ is
\begin{align}
J_y(x) = -ev_F N_F T\sum_{|\omega_n|<\omega_c}
(-i\pi)\langle \sin(\theta_{\vec{k}})g(\theta_{\vec{k}},x;\omega_n)\rangle.
\label{eq:J}
\end{align}
The spontaneous current, $J_y(x)$, gives rise to a local field, $B_z(x)$,  which can be calculated from the Maxwell equation:
\begin{align}
\frac{d B_z(x)}{dx} &= -\mu J_y(x),
\label{eq:A}
\end{align}
where the permeability, $\mu$, is related to the penetration depth $\lambda_L = \sqrt{m/e^2\mu n}$, $n$ is the normal state electron density and $m$ ($-e$) is the electron mass (charge).
To include Meissner screening in a self-consistent manner, we solve the Eilenberger equation together with the above Maxwell equation simultaneously.

Lastly, we consider the effect of surface roughness modeled by adding a disorder induced self-energy, $\hat{\Sigma}$, to $\hat{H}$ in Eq.~\eqref{eq:Eilenberger}.
Then $\hat{\Sigma}$ can be calculated within the self-consistent Born approximation from the Green's function,
\begin{align}
\hat{\Sigma}(x; \omega_n) =
\frac{i}{2\tau(x)} \langle \hat{g}(\theta_{\vec{k}},x;\omega_n)\rangle.
\label{eq:Sigma}
\end{align}
Here $\tau(x)$ is the local $x$-dependent mean free time. As a model of roughness near the surface, we take $1/\tau(x)$ to be maximum at $x=0$ and to decay to zero into the bulk.
Note that $\hat{\Sigma}(x;\omega_n)$ does not depend on the angle $\theta_{\vec{k}}$, which is a consequence of the assumption that locally the disorder scattering is isotropic.

We solve the above coupled equations for the Riccati parameters, $\Delta_1$, $\Delta_2$, $ A_y$ and $\hat{\Sigma}$ simultaneously by iteration until a stable self-consistent solution for all parameters is achieved.

\section{Edge Currents Without Meissner Screening and surface roughness}
\label{section:A0}
We first consider the edge currents without Meissner screening and surface roughness and focus primarily on the finite temperature results.

The spatial profiles of the $T=0$ edge currents are similar to those obtained in Ref.~\onlinecite{Suzuki2016}, although not identical because of the finite system size in Ref.~\onlinecite{Suzuki2016},
and can be found in Appendix~\ref{section:A0_x}.
From Fig.~\ref{fig:A0_DJM} (e, f) we see that the edge current $J_y(x)$ is finite for the chiral $d$- and $f$-wave pairings, although the integrated current is zero.
At first glance, this seems to contradict  the weak coupling GL result\cite{Huang2014}，
\begin{equation}
J_y(x) \propto k_3 \; (\Delta_2 \partial_x \Delta_1^* -c.c.)  - k_4 \; (\Delta_1^* \partial_x \Delta_2 -c.c.),
\label{eq:A0_GL}
\end{equation}
where $k_3 = k_4 = 0$ for all non-$p$-wave chiral pairing.
However, there is no contradiction since Eq.~\eqref{eq:A0_GL} only accounts for the lowest order contribution in the GL expansion.
Higher order terms, such as $\Delta_1^{*} \partial_x^3 \Delta_2-c.c.$, can lead to a small current density.
(Note that Eq.~\eqref{eq:A0_GL} implies a current along $y$ when the order parameter,
$\Delta(\theta_{\vec{k}},x)$, has a spatial phase variation along $x$. This transverse response results from the two-component chiral nature of the order parameter, as discussed
in detail in Refs.~\onlinecite{Sigrist1991,Huang2015}.)
\begin{figure}[tp]
\centering
\includegraphics[width=\columnwidth]{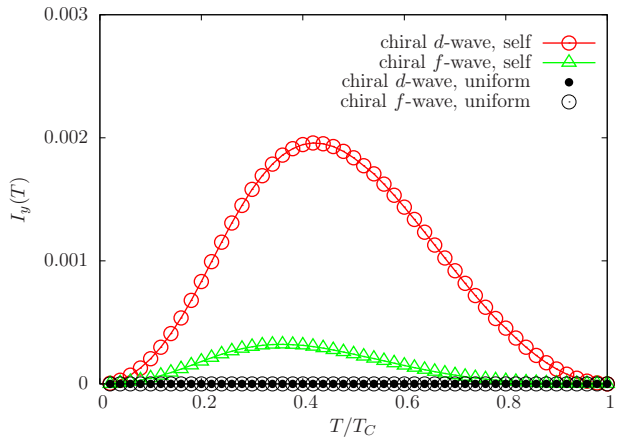}
\caption[Temperature dependence of $I_y$ for higher chirality superconductors]{Temperature dependence of the integrated edge current, $I_y(T)$, for chiral $d$- (red open circles) and $f$- (green triangles) wave with the self-consistently determined superconducting order parameter.
The black dots (open circles) are numerical results for chiral $d$ ($f$) - wave with a uniform order parameter $\Delta_1(x)=\Delta_2(x)\equiv \Delta(\mathrm{bulk})$.
$I_{y}$ is scaled by ${J}_0\xi_0$, where $J_0 = ev_F N_F T_c $ and $\xi_0 = v_F/\pi\Delta(bulk)$ is the zero temperature coherence length.}
\label{fig:A0_IyTd}
\end{figure}

At finite $T$, the total integrated current (or more precisely, the current per unit length along the $z$-direction for the quasi-$2d$ system),
$I_y\equiv \int_0^\infty J_y(x) dx$, for a chiral $p$-wave superconductor decreases monotonically with $T$ and vanishes at the superconducting transition temperature $T_c $~\cite{Kita1998, Sauls2011}.
The temperature dependence for chiral $d$- and $f$-wave superconductors is quite different, as shown in Fig.~\ref{fig:A0_IyTd}.
Although $I_y(T)=0$ at both $T=0$ and $T=T_c$, it is nonzero at $0<T<T_c$ and reaches its maximum just below $T=T_c/2$.
By contrast, as found in Ref.~\onlinecite{Huang2014}, for a uniform superconducting order parameter, i.e. $\Delta_{1}(x)=\Delta_{2}(x)\equiv\Delta(\mathrm{bulk})$, $I_y(T)\equiv 0$ for any $T$.
This result can be derived from an analytical treatment of the Eilenberger equation (see Appendix~\ref{sec:App_Eilenberger}) and a semi-classical BdG analysis.
The spatially varying order parameter is crucial for the nonzero $I_y$ at finite $T$.

\begin{figure}[tp]
\centering
\includegraphics[width=\columnwidth]{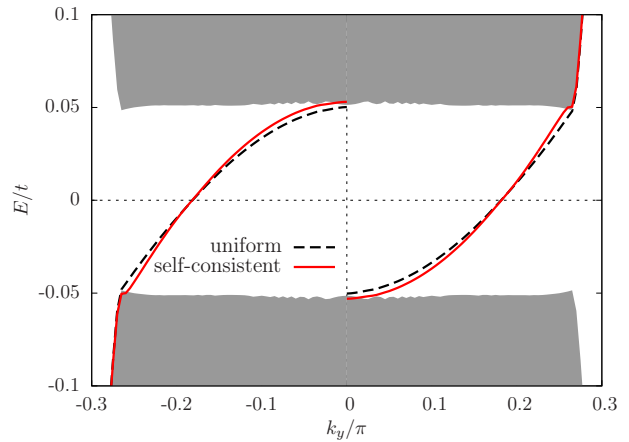}
\caption[Edge  dispersions of chiral $d$-wave models]{Edge state energy dispersion for chiral $d$-wave pairing obtained from BdG calculations with a self-consistently determined superconducting order parameter (solid red lines) and a uniform order parameter (dashed black lines).
The shaded regimes represent dense bulk energy spectra, whose details are not shown here. Inside the bulk superconducting gap there are two edge state energy dispersions crossing $E=0$
at $k_y=\pm k_F/\sqrt{2}$. }
\label{fig:A0_Ekd}
\end{figure}

To understand the above results, we first consider the case of a uniform superconducting order parameter.
In this case the edge state dispersion can be obtained from a semiclassical
BdG analysis, as given in Refs.~\onlinecite{Horovitz2003, Huang2014}. It is determined by the enhanced component of the order parameter and for our
geometry given by the following piecewise function,~\cite{Huang2014}
\begin{gather}\label{eq:Ejthetak}
E^{(j)}(\theta_{\vec{k}})=
\begin{cases}
(-1)^{(j-1)} \Delta_0\cos(m \theta_{\vec{k}}), & \text{if } m=\mathrm{even}, \\
(-1)^{j} \Delta_0\sin(m \theta_{\vec{k}}), & \text{if } m=\mathrm{odd},
\end{cases}
\end{gather}
for $-\pi/2 +(j-1)\pi/m\le \theta_{\vec{k}}< -\pi/2 +j \pi/m$, where
the edge state branch number $j=1,2,\cdots, m$.
For the chiral $d$-wave case there are two edge state branches (see Fig.~\ref{fig:A0_Ekd}) and their dispersions, in terms of $k_y=k_F\sin\theta_{\vec{k}}$, are
$E^{(j)}(k_y)=\pm \Delta_0(k_F^2-2k_y^2)/k_F^2$. At $T=0$ only the states with $E^{(j)}(\theta_{\vec{k}})\le 0$ are occupied and their contribution to the edge
current is~\cite{Huang2014}
\begin{subequations}
\begin{align}
& I_y(T=0)  \propto \sum_{j=1}^m \int \Theta(-E^{(j)}(\theta_{\vec{k}}))\;k_y \; d k_y \label{eq:Iya} \\
 & \propto \int \Theta(-E^{(1)}(\theta_{\vec{k}})) \bigg\{ \sum_{j=1}^m \sin(2\theta_{\vec{k}}+(j-1)\frac{2\pi}{m}) \bigg\} d \theta_{\vec{k}} \label{eq:Iyb} \\
 & =0,
\end{align}
\end{subequations}
where $\Theta(x)$ is the Heaviside step function and from the first line to the second we have used the periodicity of $E^{(j)}(\theta_{\vec{k}})$,
$E^{(j)}(\theta_{\vec{k}})=E^{(j+1)}(\theta_{\vec{k}}+\pi/m)$.

The last equality comes from the fact that the $\big\{ \cdots \}$ factor in Eq.~\eqref{eq:Iyb} vanishes identically for any $|m|\ne 1$.
Hence the zero $I_y(T=0)$ is a consequence of the exact cancellation between the $m$ branch contributions. Notice that the cancellation is between $m$ different $k_y$ states,
one from each of the $m$ edge state branches, for any allowed energy $E$ and it is independent of the zero temperature occupation number $\Theta(x)$.
At finite $T$, $I_y(T)$ is still given by the above integral in Eq.~\eqref{eq:Iya} and~\eqref{eq:Iyb} except
that $\Theta(x)$ is replaced by the Fermi-Dirac distribution, $n_F(x)=1/(e^{x/T}+1)$, and the gap magnitude is $T$
dependent. Since the factor $\{\cdots\}\equiv 0$ in Eq.~\eqref{eq:Iyb} is independent of $T$ we reach the conclusion that $I_y(T)\equiv 0$
for any $|m|\ne 1$, if the order parameter is uniform, in agreement with the results in Fig.~\ref{fig:A0_IyTd}. We should emphasize that the edge currents not only come from edge states but also from bulk scattering states. In the following, for qualitative understanding,
we only focus on the edge state contributions. However, in Appendix~\ref{sec:App_Eilenberger}, we show that if the order parameter is uniform, the bulk contribution
of $I_y(T)$ also vanishes at all $T$.

Next we consider the case with $\Delta_{1}$ and $\Delta_{2}$ determined self-consistently.
When the $x$-dependence of the superconducting order parameter near the surface is taken into account, $E^{(j)}(\theta_{\vec{k}})$
is no longer given by Eq.~\eqref{eq:Ejthetak} and $I_y(T=0)$ can not be expressed as Eq.~\eqref{eq:Iyb}. However, the $m$ integrals in Eq.~\eqref{eq:Iya}
still cancel at $T=0$ because they only depend on the lower and upper $k_y$ limits of each integral but not the details of $E^{(j)}(\theta_{\vec{k}})$.
These $k_y$ values remain the same as in
the uniform order parameter case and only depend on $k_F$ or $m$, because the lower $k_y$ limit is determined by the starting $\theta_{\vec{k}}$ point of each branch
dispersion, $E^{(j)}(\theta_{\vec{k}})$, while the upper limit by
$E^{(j)}(\theta_{\vec{k}})=0$.
\footnote{At these $k_y$ (or $\theta_{\vec{k}}$) points, edge states for positive chirality are degenerate in energy with those of negative chirality,
which occurs when one component of the order parameter, $\Delta(\theta_{\vec{k}},x)=\Delta_1(x) \cos(m\theta_{\vec{k}})+\Delta_2(x) \sin(m\theta_{\vec{k}})$, vanishes
for a general $x$. Therefore, the degenerate $k_y$ (or $\theta_{\vec{k}}$) points are completely fixed by the pairing symmetry of the two components,
either $\cos(m\theta_{\vec{k}})=0$ or $\sin(m\theta_{\vec{k}})=0$, regardless of whether $\Delta_1$ and $\Delta_2$ are uniform or self-consistently determined.}
This is also confirmed for the chiral $d$-wave case by self-consistent BdG, as shown in Fig.~\ref{fig:A0_Ekd}. As a consequence, from Eq.~\eqref{eq:Iya}, $I_y(T=0)=0$ remains. This result is consistent with Ref.~\onlinecite{Huang2014} and ~\onlinecite{Tada2015},
where $I_y(T=0)$ was shown to be of order $\mathcal{O}(\Delta/E_F)$ for chiral $d$- or $f$-wave; in the semiclassical approximation,
 $\Delta/E_F\rightarrow 0$ (implicit in the Eilenberger formalism), and, consequently, $I_y(T=0)=0$. 

At finite $T$, edge states with $E^{(j)}(\theta_{\vec{k}})>0$ also contribute to the current due to thermal population and the entire 
$E^{(j)}(\theta_{\vec{k}})$ dispersion matters.
Since $E^{(j)}(\theta_{\vec{k}})$ is no longer given by Eq.~\eqref{eq:Ejthetak} and $I_y(T)$ can not be written in the form of Eq.~\eqref{eq:Iyb},
the exact cancellation between the $m$ branches breaks down and gives rise to the nonzero $I_y$ at finite $T$ in Fig.~\ref{fig:A0_IyTd}.
As $T\rightarrow T_c$ one approaches equal occupation of all edge states which results in zero current.
The competition between the two factors, the imbalance between the $m$ different edge state branches and the thermal degradation of currents as one approaches $T_c$, results in the $I_y(T)$ peak around $T=T_c/2$ in Fig.~\ref{fig:A0_IyTd}.
These results could have implications for future experiments on possible higher chirality superconductors, as discussed in the conclusions.

We also note that in Fig.~\ref{fig:A0_IyTd}, as $T\rightarrow T_c$, $I_y(T)$ vanishes faster than $I_y(T)\propto T_c -T$, in stark contrast to
the chiral $p$-wave case~\cite{Kita1998,Sauls2011}. The difference comes from the fact that the lowest order nonzero contribution to the edge current density for higher chirality superconductors comes from terms, which involve higher order spatial derivatives than those in Eq.~\eqref{eq:A0_GL} for
chiral $p$-wave. For example, for chiral $d$-wave, two of these terms are $\Delta_1^* \partial_x^3 \Delta_2 -c.c.$, which predict a scaling of the current density $J_y \propto \Delta(T)^2/\xi(T)^3$ for $T\lesssim T_c$.
Here $\Delta(T)$ and $\xi(T)$ is the temperature dependent gap magnitude and coherence length, respectively. This
leads to $I_y(T)\propto J_y \xi(T)\propto (T-T_c)^2$ for $T$ near $T_c$.\footnote{In Fig.~\ref{fig:A0_IyTd} the $T$ dependence of $I_y(T)$ for chiral $d$-wave may slightly deviate from the quadratic prediction
at $T\approx T_c$ due to the finite $\theta_{\vec{k}}$- and $x$-grid sizes used in numerics and the diverging $\xi(T)$ as $T\rightarrow T_c$.}

Although we have focused on higher chirality superconductors in the above, the same conclusion that the self-consistency of the order parameter does not change
the $I_y(T=0)$ but has an effect on the finite temperature $I_y(T)$ applies to the chiral $p$-wave case as well. Since, for the chiral p-wave case, $I_y(T)$ is already large
for a uniform order parameter, the self-consistency of the order parameter only changes the finite temperature result by a relatively small amount.

Finally, in a more general lattice model with anisotropy, $I_y(T=0)$ does not need to vanish for higher chiralities. The above argument breaks down, since the $k_y$ positions
of $E^{(j)}(k_y)=0$ are not protected by any bulk band topology.

\section{Meissner Screening effect on the edge currents}
\label{section:A}
\begin{figure}[tp]
\center
\includegraphics[width=1.1\columnwidth]{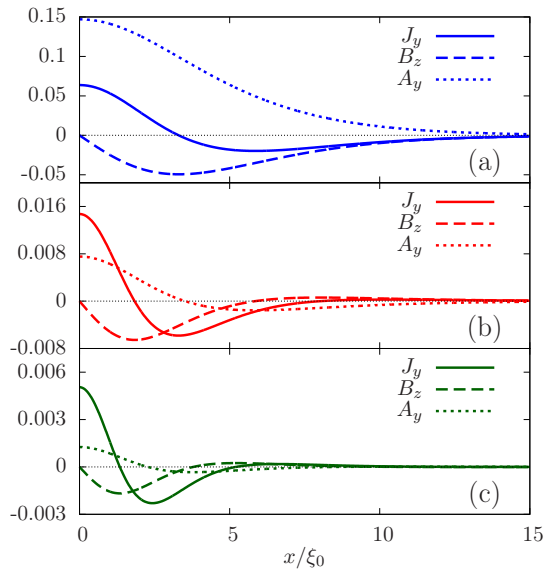}
\caption[Spatial profiles of the screened chiral edge current]{(a) - (c) Spatial dependencies of the edge current density, $J_y(x)$, induced magnetic field, $B_z(x)$, and vector potential, $A_y(x)$, with Meissner screening taken into account for chiral $p$-, $d$- and $f$-wave pairing, respectively.
GL ratio $\kappa \equiv \lambda_L/\xi_0 = 2.5$. $J_y(x)$, $A_y(x)$ and $B_z(x)$ are scaled by $J_0 = ev_F N_F T_c$, $\Delta(\mathrm{bulk})/ev_F$ and $B_c = \Phi_0/2\sqrt{2}\pi \xi_0\lambda_L$, respectively, where $\Phi_0=h/2 e$ and $T=0.02T_c$.}
\label{fig:A_JM}
\end{figure}

Meissner screening is included by solving the Eilenberger equations and the Maxwell equation simultaneously and self-consistently.
The results are shown in Fig.~\ref{fig:A_JM}. As expected, the induced magnetic field $B_z(x)$ vanishes into the bulk in all cases.
Comparing Fig.~\ref{fig:A_JM} (a) - (c) with Fig.~\ref{fig:A0_DJM} (d) - (f) (in Appendix~\ref{section:A0_x}) we see that although screening reduces the edge current magnitude by a significant fraction in the chiral $p$-wave case, the magnitude of the edge currents in the higher chirality cases is much less affected.
This is expected since the unscreened edge currents for higher chirality have contributions from different edge state branches with different
signs as well as spatial variations at different length scales (see Appendix A Fig.~8).
The resulting oscillating (with sign changes) unscreened current is, effectively, partially self-screened.
Screening also introduces one additional node in the spatial dependence of $J_y(x)$ for all chiral pairing channels due to the different length scales of the diamagnetic current ($\lambda$) and the spontaneous current ($\xi_0$).

\section{Rough Surface Effect}
\label{section:R}
We now discuss the effect of the surface roughness on the edge currents.
The surface roughness is modelled with a spatially dependent local scattering rate given by
\begin{align}
\frac{1}{\tau(x)} = \frac{1}{\tau(0)}\left(\frac{1-\tanh[(x-W)/\xi_0]}{2}\right),
\end{align}
which is maximal at $x=0$ and decays into the bulk. $W$ is the effective width of the rough regime.

\begin{figure}[tp]
\hspace*{-0.8cm}
\includegraphics[width=1.1\columnwidth]{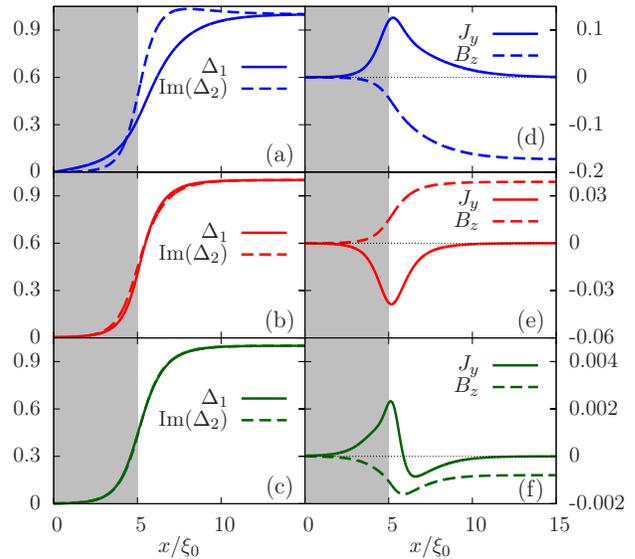}
\caption[Effects of the strong surface roughness on the unscreened edge current]{(a) - (c) Spatial dependence of $\Delta_1$ and $\mathrm{Im}(\Delta_2)$ in the presence of
surface roughness for chiral $p$-, $d$- and $f$-wave, respectively. (d) - (f) Spatial dependence of the edge current $J_y(x)$ and the
induced $B_z(x)$ in the presence of surface roughness for different chiral pairing channels. The effective rough regime with width $W=5\xi_0$ is shaded in grey.
The strength of the roughness is characterized by the shortest mean free path, $\ell_p\equiv v_F \tau(x=0)$, in the rough regime, which is $ \xi_0/\ell_p=1.0$ for
the results shown. The two order parameter components have been already scaled by their bulk values. Meissner screening is not taken into account. $T=0.02T_c$.}
\label{fig:R_A0Rstep0}
\end{figure}

To simplify the discussion, we first ignore Meissner screening.
The order parameter and edge current density computed are shown in Fig.~\ref{fig:R_A0Rstep0} for different pairing channels and strong surface roughness with a local mean free path $\ell_p\equiv v_F \tau(x=0)=\xi_0$.
The superconductivity is completely suppressed at the vacuum-superconductor interface and is nonzero in the rough regime only near $x=W$, where the surface roughness gradually disappears.

Aside from a suppression of the edge current, the most prominent feature in Fig.~\ref{fig:R_A0Rstep0}(e) is that the edge current for the chiral $d$-wave pairing case flows in a direction opposite to that of the specular surface (see Fig.~\ref{fig:A0_DJM} (e));
while the edge current direction of odd-angular momentum channels remains unaltered in the presence of the surface roughness.
The edge current inversion for the chiral $d$-wave pairing has been observed and discussed in Ref.~\onlinecite{Suzuki2016,Suzuki2017} previously.
The explanation there is that the outer edge current of the clean system, the positive part of $J_y(x)$ in Fig.~\ref{fig:A0_DJM} (e), is suppressed by surface roughness because it is closer to $x=0$, while the inner current, the negative part of $J_y(x)$ in Fig.~\ref{fig:A0_DJM} (e), survives.
The net result is then a current direction inversion. In the following, we provide an alternative explanation for the current inversion near $x=W$ for the chiral $d$-wave pairing and analyze the robustness of this effect.

\begin{figure}[tp]
\centering
\includegraphics[width=\columnwidth]{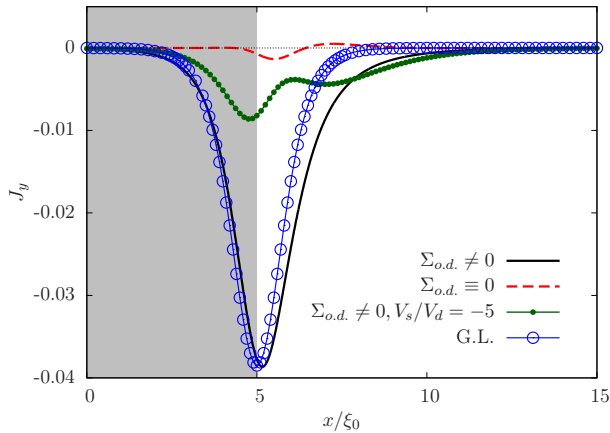}
\caption[]{Comparison of $J_y(x)$ obtained with (solid black line) and without (dashed red line) the off-diagonal impurity self energy, $\Sigma_{o. d.}$, discussed in the text.
The current from the GL free energy analysis in Eq.~\eqref{eq:JySID} (blue open circles) is plotted. Green line with dots shows that the inverted current is greatly reduced when
an $s$-wave repulsive interaction $V_s=-5 V_d$ is present. Here $V_d$ is the bulk $d$-wave attractive interaction.}
\label{fig:R_JySigmaOD}
\end{figure}

We ascribe the currents near $x=W$ in Fig.~\ref{fig:R_A0Rstep0} (e) to a disorder induced $s$-wave pairing self energy. Namely, $\hat{\Sigma}(x;\omega_n)$ defined in Eq.~\eqref{eq:Sigma} has a nonzero component off-diagonal in particle-hole space.
Let us denote it as $\Sigma_{o. d.}(x;\omega_n)$. $\Sigma_{o. d.}$ has $s$-wave symmetry and is independent of $\theta_{\vec{k}}$.
Numerically if we set $\Sigma_{o. d.}(x;\omega_n)\equiv 0$ by hand, i.e., drop the off-diagonal term in calculating $\hat{\Sigma}$ from Eq.~\eqref{eq:Sigma}, then the current near $x=W$ is almost completely suppressed, as shown in Fig.~\ref{fig:R_JySigmaOD}.

Effectively, we can interpret $\Sigma_{o. d.}$ as an additional s-wave ``order parameter", $\Delta_s$, induced by the disorder. This is possible because, although
$\Sigma_{o. d.}$ depends on the Matsubara frequency, $\omega_n$,
it is even in $\omega_n$ in the even parity $d+id$ pairing case; on the contrary, if the bulk pairing has an odd parity, then $\Sigma_{o. d.}$ is
an odd function of $\omega_n$.
The $s$-wave pairing self energy term is allowed to mix with the original order parameter, which is non-s wave, because the edge breaks inversion symmetry.

From Eq.~\eqref{eq:Sigma} we see that the $s$-wave $\Sigma_{o. d.}$ comes from the anomalous Green's function $f(\theta_{\vec{k}},x;\omega_n)$ having a non-zero $s$-wave component.
This $s$-wave component is sub-dominant to the bulk $d+i d$ pairing and induced by the suppression of the $d+id$ pairing near $x=W$,
which is in turn due to the edge and the disorder. In weak-coupling GL theory, this can be seen from the following mixed gradient term~\cite{Soininen1994,Berlinsky1995}
in free energy
\begin{gather}
+4 A_4 (D_x \Delta_s^* D_x \Delta_1 + c.c.), \label{eq:mix1}
\end{gather}
derived in Appendix~\ref{sec:App_GL}. Here $A_4>0$, $D_x\equiv \frac{v_F}{2}\partial_x$ and $\Delta_s$ is the sub-dominant s-wave order parameter induced near $x=W$.
This term favors a nonzero $\Delta_s$ where $\Delta_1$ has a spatial variation, which is most significant near $x=W$. The sign of $\Delta_s$ can be determined by
minimizing Eq.~\eqref{eq:mix1}. Given that $\Delta_1$ is real and $\partial_x \Delta_1 > 0$ near $x=W$ (see Fig.~\ref{fig:R_A0Rstep0}), $\Delta_s$ is
real as well and also $\partial_x \Delta_s < 0$, which leads to $\Delta_s>0$ since $\Delta_s=0$ in the bulk. 
In other words, the sign of $\Delta_s$ is the same as that of $\Delta_1$.

Using $\Delta_s$ we can now understand the spontaneous current $J_y(x)$ near $x=W$ in Fig.~\ref{fig:R_JySigmaOD}.
Since the spatial variations of $\Delta_1$ and $\Delta_2$ are almost identical near $x=W$, the spontaneous current due to the original $d+id$ components
is greatly suppressed.
As a consequence, the current mainly comes from the $s+i d_{xy}$ pairing components. This current can be derived from another mixed gradient term~\cite{Soininen1994,Berlinsky1995} in the GL free energy
\begin{gather}
\mathcal{F}_{\mathrm{mix}} \equiv  +A_4 \, v_F^2 \, \bigg\{ \partial_x \Delta_2 \partial_y \Delta_s^*  + \partial_y \Delta_2 \partial_x \Delta_s^* + c.c \bigg\}. \label{eq:mix2}
\end{gather}
For the half-infinite geometry, the spontaneous current is along $y$-direction.
$J_y$ can be obtained by minimal coupling $\mathcal{F}_{\mathrm{mix}}$ to the vector potential $A_y$ and taking a functional derivative of $\mathcal{F}_{\mathrm{mix}}$ with respect to $A_y$.
The result is
\begin{gather}
J_y \propto -2 e \, i\, \bigg\{ \Delta_2 \partial_x \Delta_s^* - \Delta_s^* \partial_x \Delta_2 - c.c.\bigg\}, \label{eq:JySID2}
\end{gather}
with a positive proportionality constant.
$e>0$ is the magnitude of an electron charge. To a good approximation, the spatial variation of $\Delta_s$ follows that of the local scattering rate since $\Delta_s \sim \Sigma_{o.d.} \propto 1/\tau(x)$. So we can take
\begin{gather}
\Delta_s(x)= \Delta_s \;\mathrm{sgn}(\Delta_1)\; \frac{\tau(0)}{\tau(x)}
\end{gather}
where $\Delta_s>0$ is the overall magnitude and we have made the sign dependence of $\Delta_s$ on $\mathrm{sgn}(\Delta_1)$ explicit.
Here $\Delta_1$ is the bulk value of the $d_{x^2-y^2}$ component order parameter.
The spatial variation of the $i d_{xy}$ in Fig.~\ref{fig:R_A0Rstep0} (b) can be approximated by
\begin{gather}
\Delta_2(x)= i \; \mathrm{Im}(\Delta_2) \frac{\tanh[(x-W)/\ell_h]+1}{2},
\end{gather}
where $\ell_h$ is the healing length of the $id_{xy}$ component near $x=W$ and it can be roughly taken as the maximal local mean free path: $\ell_h\approx \ell_p=\xi_0$.
Then from the expression of $J_y$ in Eq.~\eqref{eq:JySID2} we have
\begin{gather}~\label{eq:JySID}
J_y(x) = - J_0 \; \mathrm{sgn}(\Delta_1 \;\mathrm{Im}(\Delta_2)) \; \mathrm{sech}^2 \frac{x-W}{\xi_0},
\end{gather}
where $J_0>0$ is a constant that sets the maximal $|J_y(x)|$ magnitude. Using $J_0\approx 0.04$, this gives a current profile in Fig.~\ref{fig:R_JySigmaOD} (open circles) very similar to that from the numerical Eilenberger solution (black solid line).
Note that $J_y(x)$ is still odd in the chirality of the $d+id$ order parameter, as expected.
The dependence on $\mathrm{sgn}(\Delta_1)$ is inherited from the disorder induced $\Delta_s(x)$.

The GL explanation presented here, as well as the lattice BdG results, depends only on frequency-independent order parameters and their spatial derivatives,
but in the Eilenberger calculation the spontaneous current can also be related to odd-frequency pairing components of the anomalous Green's functions~\cite{Suzuki2016}.
The odd-frequency pairing appears as derivatives of an order parameter in the GL analysis after the frequency is integrated over.
So the two, GL and Eilenberger odd frequency pairing, are connected, but the GL analysis is physically more transparent.
For instance, the GL formulation shows that the current inversion depends not only on the presence of $\Delta_s$ but also on the relative phase between $\Delta_s$ and the bulk
$d+id$ order parameter components.
The phase is determined in GL by minimizing the free energy term in Eq.~\eqref{eq:mix1}.
Also, for a spontaneous edge current discussion the frequency independent GL analysis seems more natural.

The non-inversion of current in the chiral $p$-wave case (Fig.~\ref{fig:R_A0Rstep0}(d)) can also be understood within the GL framework.
In the chiral $p$-wave case, without strong surface roughness, the GL current is dominated by $J_y \propto k_3  (\Delta_2 \partial_x \Delta_1^* -c.c.) - k_4 (\Delta_1^* \partial_x \Delta_2-c.c.)$ with coefficients $k_3,k_4>0$~\cite{Huang2014}.
This remains true in the presence of strong surface disorder as disorder does not introduce any new frequency independent order parameter since $\Sigma_{o.d.}(\omega_n)$ is completely odd in frequency.
Disorder enhances the order parameter derivative term $\partial_x \Delta_1$, that also has an $s$-wave symmetry.
Due to this enhancement the magnitude of $k_3$ and $k_4$ become different ($k_3=k_4$ without surface disorder) such that $k_3 ( \Delta_2 \partial_x \Delta_1^*-c.c.)$ dominates the current.
Then with the spatial profiles of $\Delta_1$ and $\Delta_2$ given in Fig.~\ref{fig:R_A0Rstep0}(a), it is easy to see that $J_y$ remains positive in the presence of strong surface roughness,
so there is no current inversion.
The current in the chiral $f$-wave case can be understood similarly, but the analysis is more involved as order parameter derivatives higher than the first order are needed and therefore we do not elaborate on this here.

However, we should emphasize that the edge current inversion seen for chiral $d$-wave pairing in the continuum limit is not universal.
Away from the continuum limit, the direction of the current can depend on surface orientation and microscopic details.
For example, for chiral $d$-wave pairing on a triangular lattice, with the edge along the zigzag direction and the chemical potential near half filling (see Appendix~\ref{sec:App_BdG}), the current in the absence of disorder is opposite to that in the continuum limit.
In this case, there is no current inversion due to the surface disorder. Furthermore, since the current inversion requires an induced $s$-wave component, even in the continuum limit with strong edge disorder, the effect is reduced if the $s$-wave channel is repulsive.
In physical systems, unconventional pairing is usually accompanied by a sufficiently strong short range Coulomb interaction that leads to repulsive interactions in the $s$-wave channel.
In Fig.~\ref{fig:R_JySigmaOD} (green line with dots), we show the result of including $s$-wave repulsion self-consistently.

\begin{figure}[tp]
\centering
\includegraphics[width=1.1\columnwidth]{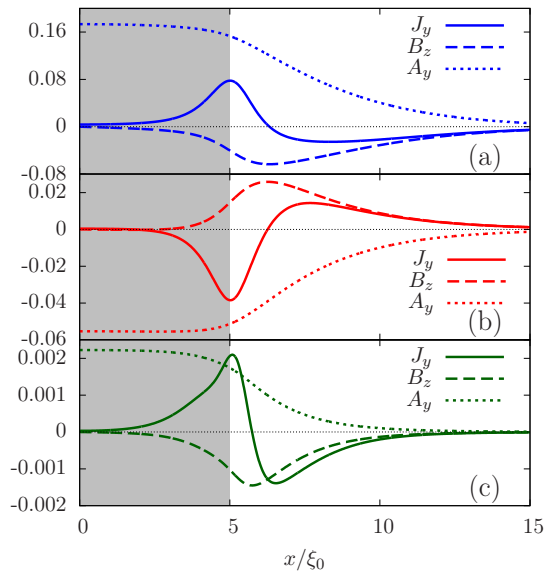}
\caption[Effects of the strong surface roughness on the screened edge current]{Spatial dependencies of $J_y(x)$,
$B_z(x)$ and $A_y(x)$ with Meissner screening for chiral $p$-, $d$- and $f$-wave pairing (from top to bottom), in the presence of a rough surface in a region of width $W=5\xi_0$ and with $\ell_p=\xi_0$.}
\label{fig:R_AR1.0}
\end{figure}

Finally we discuss the effect of Meissner screening, which so far has been neglected. Fig.~\ref{fig:R_AR1.0} shows the $J_y(x)$, $B_z(x)$ and $A_y(x)$ obtained in the presence of Meissner screening.
The major effect of the Meissner screening is to induce an additional sign change in $J_y(x)$ due to the diamagnetic current such that the total integrated current
$\int_0^{\infty} J_y(x) dx \propto B_z(x=\infty)=0$, as required by Eq.~\eqref{eq:A}.

\section{Conclusion} \label{sec:conclusion}
To summarize, we have considered the effects of finite temperature, Meissner screening, and surface roughness on the spontaneous edge current for
higher chirality superconductors in the continuum limit using the quasiclassical Eilenberger formalism.
We find that the integrated edge current for higher chirality superconductors is finite at finite $T$, although it vanishes at $T=0$~\cite{Huang2014}.
It achieves its maximum near $T=T_c/2$. The self-consistency of the superconducting order parameter was found to be crucial for understanding this temperature dependence. We also find that Meissner screening effects on the edge current are much weaker for the higher chirality superconductor,
compared with that for the chiral p-wave case.

Furthermore, we have studied the rough surface effects on the edge current by modeling the surface roughness as an effective disorder scattering.
Similar to Ref.~\onlinecite{Suzuki2016}, we have found that the edge current direction is inverted by the surface roughness in the chiral $d$-wave case.
We ascribe the inverted edge current to a disorder induced sub-dominant $s$-wave pairing ``order parameter"
in the rough surface regime and explain the current inversion using the GL analysis.

However, we find that this current inversion is not universal beyond the continuum limit and
can depend on microscopic details, such as the surface orientation and the filling level of the sample, as seen from our self-consistent lattice BdG calculations.
Furthermore, since the current inversion requires the presence of an induced $s$-wave order parameter, the effect is suppressed by any repulsion in the s-wave channel.
In general, s-wave repulsion is expected to be quite large for most unconventional superconductors. Consequently, the primary feature of edge currents in disordered
chiral $d$-wave (as well as higher chirality) superconductors is that they are expected to be quite small, relative to the analogous chiral p-wave case, and
the direction of the current is sensitive to microscopic details.

Experimentally a direct study of the edge currents has been conducted only for the chiral $p$-wave superconductor candidate material $\mathrm{Sr_2 Ru O_4}$ so far.
However, as more and more candidate materials for higher chirality superconductivity, such as $\mathrm{Sr As Pt}$, doped Graphene, $\mathrm{UPt_3}$ and $\mathrm{URu_2Si_2}$, become available, similar searches for edge currents may be undertaken.
Our results, especially the finite temperature behavior of the integrated current $I_y(T)$ and the non-universal aspect of the current inversion in the presence
of disorder in the chiral $d$-wave pairing case, could be important for understanding these materials.

\acknowledgements
We would like to thank John Berlinsky and Wen Huang for helpful discussions. This work is supported by Natural Sciences and Engineering Research Council of Canada (NSERC)
and Canadian Institute for Advanced Research (CIFAR).

\appendix \label{sec:appendix}
\begin{appendix}

\section{Spatial profile of the edge current without Meissner screening}
\label{section:A0_x}
\begin{figure}[tp]
\hspace*{-0.6cm}
\includegraphics[width=1.1\columnwidth]{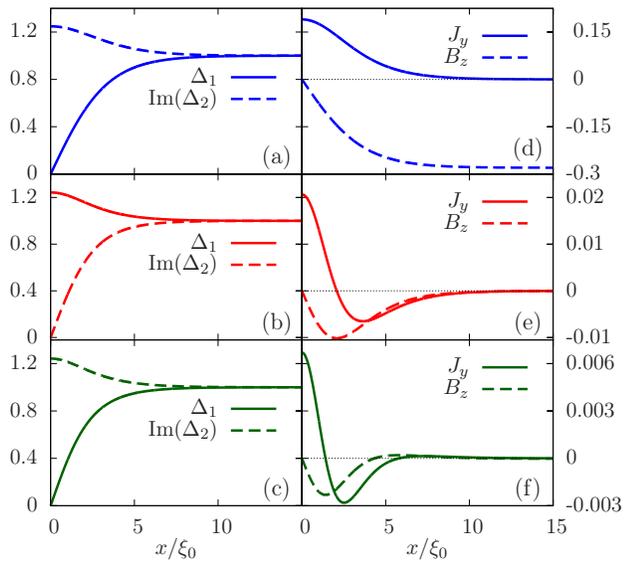}
\caption[Spatial profiles of the unscreened chiral OP]{ (a) - (c) Spatial dependence of $\Delta_1$ and $\mathrm{Im}(\Delta_2)$ with specular surface for chiral $p$-, $d$- and $f$-wave, respectively.  (d) - (f) Spatial dependence of the edge current $J_y(x)$ and the
induced $B_z(x)$ with specular surface for different chiral pairing channels. Note the different vertical scales in (d) - (f). Meissner screening is not taken into account. $T=0.02T_c$.}
\label{fig:A0_DJM}
\end{figure}

In this appendix, we show the spatial profiles of the $T=0$ spontaneous edge currents for different pairing channels.
Since at $T=0$ the number of Matsubara frequencies in the numerical calculation diverges, we use $T=0.02 T_c$ to approximate $T=0$.
Unless specified otherwise, the pairing energy cutoff is chosen to be $\omega_c=10 T_c$.
The spatial profile of the pairing components are shown in Fig.~\ref{fig:A0_DJM} (a) - (c).
In all cases, the pairing component that is odd under $k_x\rightarrow -k_x$ drops to zero at the edge while the other component (even under $x$-inversion), is enhanced near the edge\cite{Matsumoto1999}.

\begin{figure}[tp]
\centering
\includegraphics[width=\columnwidth]{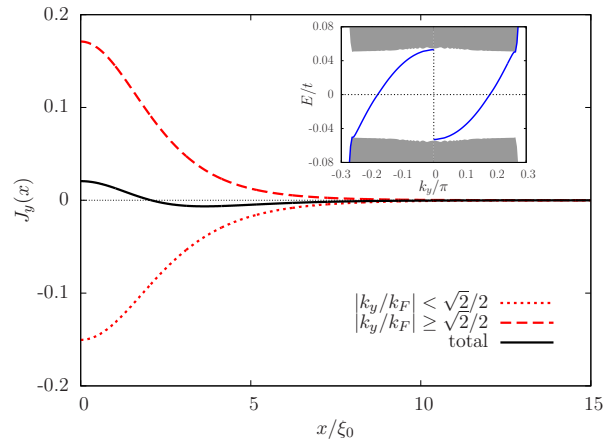}
\caption[Current channels] {Decomposition of the current density for the chiral $d$-wave pairing into different branches (i.e. different $k_y$ ranges).
Inset: edge state energy dispersion for a chiral $d$-wave superconductor obtained from a lattice BdG calculation; the horizontal axis is $k_y/\pi$; the two blue solid lines are edge state dispersions while the grey shaded regime represent the bulk state energy spectrum.}
\label{fig:A0_channel}
\end{figure}

Fig.~\ref{fig:A0_DJM} (d) - (f) shows the spatial profile of the spontaneous edge current density and the induced local magnetic field.
For chiral $m$-wave pairing, the current density changes sign $|m|-1$ times along the $x$-direction
(the second sign change for the chiral $f$-wave can not be resolved in Fig.~\ref{fig:A0_DJM} (f) because the current magnitude is too small).
This results from the $|m|$ branches of edge states carrying the edge current with different signs and different length scales (see Fig.~\ref{fig:A0_channel}
for the chiral $d$-wave pairing, for example).
Since chiral $p$-wave has a single edge mode, its edge current does not change sign and the integrated edge current can be sizable;
while for higher chirality, the integrated edge current is negligible due to the multiple sign changes and vanishes at $T=0$. This fact has been emphasized in previous studies~\cite{Huang2014, Tada2015}.
The edge currents are carried not only by the chiral Majorana edge modes but also by the bulk scattering states, which partially cancel the edge mode current~\cite{Stone2004}.
However, for a qualitative understanding of the edge current, one often can focus on the edge mode contributions only.

Fig.~\ref{fig:A0_DJM} (d) - (f) also shows that the bulk magnetic field induced by the edge current vanishes at $T=0$ for chiral $d$- and $f$-wave. This is consistent with the total current integrating to zero at $T=0$.

\section{Integrated edge currents for a uniform superconducting order parameter and no screening}
\label{sec:App_Eilenberger}
For a uniform superconducting order parameter and no vector potential
$\vec{A}$, the Eilenberger equation can be solved analytically~\cite{Matsumoto1995,Matsumoto1995a,Matsumoto1996,Matsumoto1999,Sauls2011,Huang2015}.
Decomposing the quasi-classical Green's function matrix $\hat{g}$ in terms of Nambu particle-hole Pauli matrices we can write
$\hat{g} = g_1\hat{\tau}_1+g_2\hat{\tau}_2+g_3\hat{\tau}_3$, with~\cite{Matsumoto1995a,Sauls2011,Huang2015}
\begin{align}
g_3(\theta_{\vec{k}}, x; \omega_n) =& \frac{\omega_n}{\lambda}
+  \frac{\Delta_1}{\lambda} \frac{\omega_n\Delta_1 - i s\lambda\Delta_2}{\omega_n^2 +\Delta_2^2 }
e^{-2\frac{\lambda}{|v_{F_x}|}x},
\label{eq:App_g3}
\end{align}
where, $s \equiv \mathrm{sgn}(v_{F_x})=\mathrm{sgn}(\cos \theta_{\vec{k}})$ and $\lambda=\sqrt{\omega_n^2+\Delta^{2}}$.
$\Delta_1\equiv \Delta_1(\theta_{\vec{k}})$ and $\Delta_2\equiv \Delta_2(\theta_{\vec{k}})$ are the vanishing and enhanced superconducting order parameter component, respectively.
For example, for chiral $p$-wave, $\Delta(\theta_{\vec{k}}) =\Delta(\cos \theta_{\vec{k}} + i\sin \theta_{\vec{k}})$, $\Delta_1=\Delta \cos \theta_{\vec{k}}$ and $\Delta_2=\Delta \sin \theta_{\vec{k}}$;
while for chiral $d$-wave, $\Delta(\theta_{\vec{k}}) =\Delta(\cos 2\theta_{\vec{k}} + i\sin 2\theta_{\vec{k}})$,  $\Delta_1=\Delta \sin 2 \theta_{\vec{k}}$ while
$\Delta_2=\Delta \cos 2 \theta_{\vec{k}}$. Note that the definitions of $\Delta_1$ and $\Delta_2$ here are different from those used elsewhere in the paper.

The local current density $J_y(x,T)$ can be computed from Eq.~\eqref{eq:J} with the energy cutoff $\omega_c$ sent to infinity and
the integrated current is $I_y(T)=\int_0^\infty J_y(x,T)dx$.
In Ref.~\onlinecite{Huang2014}, it was shown that $I_y(T=0)=0$ for any non-chiral-p-wave pairing. Here we give the expression of $I_y(T)$.
This finite $T$ expression of $I_y(T)$ has been derived for the chiral $p$-wave pairing in Ref.~\onlinecite{Sauls2011}.
Our derivations parallel those and we only give the final result here:
\begin{align}
\frac{I_{y}(T)}{e N_F v_F^2/8}
 &= 2 \bigg\langle \frac{v_{F_x}}{v_F} \frac{v_{F_y}}{v_F} \Delta_{1}\Delta_{2} \bigg\{ \pi \frac{\tanh{\left(\frac{|\Delta_{2}|}{2T}\right)}}{|\Delta_{1}||\Delta_{2}|} \nonumber \\
 & -  2 \int_{0}^{\infty}dy\frac{\tanh{\left( \frac{|\Delta|}{2T} \cosh{y} \right)}}{\Delta_{2}^{2}\sinh^{2}{y}+\Delta_{1}^{2}\cosh^{2}{y}}\bigg\} \bigg\rangle_{\theta_{\vec{k}}}.
\label{eq:A0T_ITana}
\end{align}
Inside the $\big\{\cdots \big\}$ the first term comes from a complex contour integral around the pole on the complex $\omega_n$ plane at $\omega_n=i|\Delta_2|$, and represents the edge mode contribution;
while the second term originates from the branch cut on the complex $\omega_n$ plane running from $\omega_n=i|\Delta|$ to $\omega_n=i \infty$.
The branch cut contribution comes from the bulk scattering states with quasiparticle energies $\ge|\Delta|$.
In general, the two contributions can both be nonzero.

In Eq.~\eqref{eq:A0T_ITana}, whether $I_y(T)=0$ or not is solely determined by the rotational symmetry of the integrand with respect to $\theta_{\vec{k}}$.
For pairing with $\Delta_{\vec{k}}= \Delta ( \cos{m\theta_{\vec{k}}}+i\sin{m\theta_{\vec{k}}})$, in the integrand of Eq.~\eqref{eq:A0T_ITana}, the combination of $\Delta_1\Delta_2 \big\{ \cdots \big\}$ is invariant under $2|m|-$fold rotation of $\theta_{\vec{k}}$;
on the other hand, the velocity product $v_{F_x} v_{F_y}$ is $2$-fold rotation symmetric.
Hence, the entire integrand can be decomposed into a sum of two terms which are invariant under either $2 |m|+2$ or $2|m|-2$ fold $\theta_k$ rotation.
Since the integral vanishes as long as $2 |m|+2 \ne 0$ and  $2|m|-2 \ne 0$, we conclude that, for any $|m|\ne 1$, $I_y(T)\equiv 0$ at any finite $T$, for the case of a uniform superconducting order parameter.

Although the above results in this section are derived for a uniform order parameter, we note that, at $T=0$, the bulk scattering state contribution of $I_y(T=0)$,
$I_y^{\mathrm{bulk}}$, remains zero even when the order parameter is self-consistently determined. This is because $I_y^{\mathrm{bulk}}$ can be written as
$I_y^{\mathrm{bulk}}\propto \int_{-\pi/2}^{\pi/2} Q_m(\theta_{\vec{k}}) \sin \theta_{\vec{k}} \cos \theta_{\vec{k}} d \theta_{\vec{k}}$~\cite{Huang2014}, where
the factor $\sin \theta_{\vec{k}} \cos \theta_{\vec{k}} d \theta_{\vec{k}}$ comes from $k_y d k_y$ and $Q_m(\theta_{\vec{k}})$ is the accumulated charge near the surface
due to the phase shift of all filled bulk scattering states at $\theta_{\vec{k}}$. As shown in Refs.~\onlinecite{Stone2004} and \onlinecite{Goldstone1981}, 
$Q_m(\theta_{\vec{k}})$ only depends on the asymptotic phase of the order parameter in the bulk, which is unaffected by the presence of the surface. Therefore, 
$I_y^{\mathrm{bulk}}(T=0)=0$ remains even when the spatial variations of the order parameter near the surface is taken into account. 

\section{Self-consistent BdG calculation of edge current with surface roughness for chiral $d$-wave pairing}
\label{sec:App_BdG}
Here, we show that the current inversion due to the surface roughness seen in the continuum limit for the chiral $d$-wave pairing is non-universal and is not always present when lattice effects are included.
The existence of the current inversion depends on microscopic details, such as the edge orientation, band structure and carrier doping levels.

We consider a two-dimensional triangular lattice with the following BdG Hamiltonian
\begin{align}
H = \sum_{\langle \vec{r},\vec{r}^\prime \rangle} \left[-t  c^{\dagger}_{\vec{r}}c_{\vec{r}^\prime}
+ \Delta_{\vec{r},\vec{r}^\prime}c^{\dagger}_{\vec{r}^\prime}c^{\dagger}_{\vec{r}} + h.c. \right] - \sum_{\vec{r}} \mu_{\vec{r}} c^{\dagger}_{\vec{r}}c_{\vec{r}}  ,
\end{align}
where $\langle \vec{r},\vec{r}^\prime\rangle$ means only nearest neighbor (NN) hopping, $t$, and pairing, $\Delta_{\vec{r},\vec{r}^{\prime}}$, are considered and $\mu_{\vec{r}}$ is the local chemical potential.
For the clean system without edges, the normal state energy dispersion is given by $\epsilon_{\vec{k}}=-2t\left[\cos{k_x}+2\cos(\sqrt{3}k_y/2)\cos{(k_x/2)}\right]$,
where the lattice spacing is set to unity. We have chosen the $x-$direction along one of the three lattice bond directions.

The chiral $d$-wave superconducting order parameter is defined on each NN bond, $\frac{\vec{r}+\vec{r}^\prime}{2}$, as $\Delta_{\vec{r},\vec{r}^\prime}=\Delta(\frac{\vec{r}+\vec{r}^\prime}{2})
e^{i\, 2 \, \mathrm{Arg}[(x^\prime-x)+i(y^\prime -y)] }$. Without edges and disorder, $\Delta(\frac{\vec{r}+\vec{r}^\prime}{2})\equiv \Delta_0$ (a constant), and the order parameter in $\vec{k}$ space is
$\Delta_k = \Delta_0\left[\cos{k_x}-\cos(\sqrt{3}k_y/2)\cos{(k_x/2)}\right]+i\Delta_0\sqrt{3}\sin(\sqrt{3}k_y/2)\sin(k_x/2)$.
In the presence of edges, the order parameter magnitude $\Delta(\frac{\vec{r}+\vec{r}^\prime}{2})$ becomes position dependent;
however, we keep the phase $e^{i\, 2 \, \mathrm{Arg}[(x^\prime-x)+i(y^\prime -y)] }$ the same to ensure the pairing is chiral $d$-wave.
$\Delta_{\vec{r},\vec{r}^\prime}$ is determined self-consistently within BdG (details can be found in Refs.~\onlinecite{Lederer2014, Huang2015}).

Surface roughness is moddelled by adding a random impurity potential $V^{\text{imp}}_{\vec{r}}$ ($\mu_{\vec{r}} =\mu+V^{\mathrm{imp}}_{\vec{r}}$) to sites within a width $W$ of the edge.
The impurity density in the rough regime is $n_{\mathrm{imp}}=0.2$ per lattice site and $V^{\text{imp}}_{\vec{r}}$ is uniformly distributed in the range $[-V^{\text{imp}}, V^{\text{imp}}]$.
The current calculated is averaged over different impurity configurations.

\begin{figure}[tp]
\center
\includegraphics[width=1.2\columnwidth]{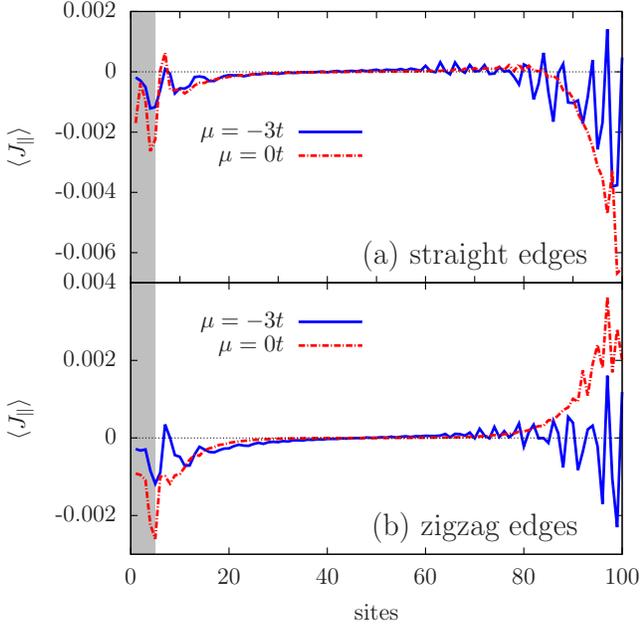}
\caption[BdG_zizzag] {Edge currents for the chiral d-wave pairing obtained from self-consistent BdG on a triangular lattice with different edge directions and chemical potentials.
The shaded regime on the left has surface roughness; while the right surface is specular without disorder.
We choose the rough regime width to be $W=5$ lattice sites.
A relatively larger temperature $T=0.1T_c$ has been chosen to reduce the Friedel oscillations in the current.
The impurity potential strength, $V^{\mathrm{imp}}=15t$, and the impurity density, $n_{\mathrm{imp}} = 0.2$ per site, are large such that the effective local mean free path is short, same as in the Eilenberger calculation, where current reversal is seen.}
\label{fig:BdG_J}
\end{figure}

We consider two different types of edges of the triangular lattice, straight and zigzag, and use periodic boundary conditions for the direction parallel to the edges. The current along each type of edge, denoted as $J_{\parallel}$,
is calculated for two different filling levels, $\mu=0$ (half-filling) and $\mu=-3t$; the results are shown in Fig.~\ref{fig:BdG_J}.
For straight edges, there is an edge current inversion due to the surface roughness at both $\mu=0t$ and $\mu=-3t$;
while for zigzag edges, the current inversion is seen only at $\mu=-3 t$, not at $\mu=0$.
In the specular surface (right edge of Fig.~\ref{fig:BdG_J}) case, the $\mu=0$ edge current of the straight edge and that of the zigzag edge flow in opposite directions. Similar results have been observed for chiral $p$-wave pairing in Ref.~\onlinecite{Bouhon2014}.
Consequently, the direction of the current for an ideal edge and the presence of current inversion due to disorder both are sensitive to microscopic details.

\section{Ginzburg-Landau analysis of $d_{x^2-y^2}+i d_{xy}$ with sub-dominant $s$-wave pairing}
\label{sec:App_GL}
To understand the role in current reversal of sub-dominant $s$-wave pairing $\Delta_s$ induced near the surface of the $d_{x^2-y^2}+i d_{xy}$ superconductor,
we derive the GL free energy from the anomalous Green's functions, $f$ and $\bar{f}$, obtained from the Eilenberger equations.
For the half-infinite plane, the derivation closely follows that for $p_x+i p_y$ pairing in Ref.~\onlinecite{Matsumoto1999}.
Hence, in the following, we only give key steps that are different from Ref.~\onlinecite{Matsumoto1999}.

The GL free energy is an expansion in terms of
\begin{gather}
\frac{|\Delta|}{T}\equiv \frac{\mathrm{max}\big\{ |\Delta_s|,|\Delta_1|,|\Delta_2|\big\}}{T},\ \ \ \ \frac{D}{T}\equiv\frac{|v_{Fx}|\partial_x}{T},
\end{gather}
near $T=T_c$, where both ratios are small.
However, we expect qualitative features, such as the relative phase of $\Delta_s$, to survive at low temperature.
$f$ and $\bar{f}$, can be expanded in powers of $|\Delta|/T$ and $D/T$ and has been done up to the fifth order in Eq.~(A.6)
of Ref.~\onlinecite{Matsumoto1999}. The derivation here becomes different starting at the form of the order parameter
\begin{align}
 \Delta & = \Delta_s + \Delta_1 \cos 2\theta_{\vec{k}} + \Delta_2 \sin 2\theta_{\vec{k}},
\end{align}
where $\Delta_s,\Delta_1$, and $\Delta_2$ are the $\theta_{\vec{k}}$ independent parts of the $s$, $d_{x^2-y^2}$, and $id_{xy}$ order parameters.
They are complex and spatially dependent. They satisfy the following BCS gap equations
\begin{widetext}
\begin{gather}
   \begin{pmatrix}
    \Delta_s \\
    \Delta_1 \\
    \Delta_2
  \end{pmatrix}
  =
  \pi T \sum_{0<\omega_n < \omega_c} \int_{-\pi}^{\pi} \frac{d\theta_{\vec{k}}}{2\pi}
  \begin{pmatrix}
     V_s \\
    2 V_d \cos 2\theta_{\vec{k}} \\
    2 V_d \sin 2\theta_{\vec{k}}
  \end{pmatrix}
  \bigg[ f(\theta_{\vec{k}},x;\omega_n)+\bar{f}^*(\theta_{\vec{k}},x;\omega_n) \bigg]. \label{eq:App_GL_BCS}
\end{gather}
\end{widetext}
where $V_s>0$ and $V_d>0$ are the attractive interactions for $s$ and $d+id$, respectively.
Substituting the expressions of $f$ and $\bar{f}$ from Eq.~(A.6) of Ref.~\onlinecite{Matsumoto1999} into the above BCS gap equations, we obtain three coupled equations for $\Delta_s,\Delta_1$ and $\Delta_2$
up to third order in the total power of $|\Delta|/T$ and $D/T$. These equations should be reproduced by the GL free energy $\mathcal{F}$ through
a Euler-Lagrange equation~\cite{Matsumoto1999}.

Omitting derivation details, the final results for $\mathcal{F}$ are $\mathcal{F} \propto \mathcal{F}_2 + \mathcal{F}_4$ with
\begin{widetext}
\begin{subequations}
\begin{align}\label{eq:mathcalF}
  \mathcal{F}_4 & = + 4 A_4 \bigg\{  |\Delta_s|^4 + \frac{3}{8}(|\Delta_1|^4+|\Delta_2|^4) +2 |\Delta_s|^2(|\Delta_1|^2+|\Delta_2|^2\big) + \frac{1}{2}|\Delta_1|^2 |\Delta_2|^2 \nonumber \\
  & \qquad \qquad  + \frac{1}{2} (\Delta_s^*)^2 \big[\Delta_1^2 +\Delta_2^2\big] +
  \frac{1}{2} \Delta_s^2 \big[(\Delta_1^*)^2 +(\Delta_2^*)^2\big]  + \frac{1}{8} \big[\Delta_2^2 (\Delta_1^*)^2 +(\Delta_2^*)^2 \Delta_1^2 \big]  \nonumber \\
  & \qquad \qquad  +2 |D_x \Delta_s|^2 +  |D_x \Delta_1|^2 + |D_x \Delta_2|^2
   + \bigg(D_x \Delta_s^* D_x \Delta_1 + D_x \Delta_s D_x \Delta_1^* \bigg) \bigg\},
\end{align}
\end{subequations}
\end{widetext}
with $D_x\equiv \frac{v_F}{2}\partial_x$ and $A_4 \equiv \frac{1}{4} \frac{2^3-1}{2^3} \frac{\zeta(3)}{(\pi T)^2}$.
Here $\zeta(z)$ is the Riemann zeta function. $\mathcal{F}_2$ is the second order term which is not shown as it is irrelevant to
our discussion.
In $\mathcal{F}_4$, the last term involves mixed gradients of $\Delta_s$ and $\Delta_1$.
This term has been discussed in Refs.~\onlinecite{Soininen1994,Berlinsky1995,Wang1999}.
This term can induce a nonzero $\Delta_s$ where $\partial_x \Delta_1 \ne 0$ and determine the phase of $\Delta_s$ relative to $\Delta_1$ and $\Delta_2$.
In addition to the mixed gradient term there are other terms in $\mathcal{F}_4$, such as $\Delta_s^2(\Delta_1^*)^2 + c.c.$, that can also affect the phase of $\Delta_s$.
However, they are higher order for a small, spatially varying $\Delta_s$.

There are additional mixed gradients terms that are absent for a $y$-translational invariant system.
However they enter in the current and can be obtained from the one in $\mathcal{F}_4$ by four-fold rotation symmetries or can be derived as in Ref.~\onlinecite{Wang1999}.
From Ref.~\onlinecite{Wang1999}, these additional mixed gradient terms are
\begin{gather}
  - 4 A_4 \bigg(D_y \Delta_s^* D_y \Delta_1 + D_y \Delta_s D_y \Delta_1^* \bigg) \nonumber \\
  + 4 A_4 \bigg(D_x \Delta_s^* D_y \Delta_2 + D_y \Delta_s^* D_x \Delta_2 + c.c. \bigg).
\end{gather}
The important terms for the spontaneous current discussion in the main text are those from the second line. Other mixed gradient terms do not contribute a spontaneous current along the $y$-direction for
the half-infinite geometry that we considered.

\end{appendix}


%

\end{document}